\documentclass[a4paper,10pt]{article}

\usepackage{amsfonts,amssymb,amsmath,xcolor}
\usepackage[ruled,noline]{algorithm2e} 
\usepackage{verbatim}
\usepackage{hyperref}

\newcommand{\qed}{\nobreak \ifvmode \relax \else
      \ifdim\lastskip<1.5em \hskip-\lastskip
      \hskip1.5em plus0em minus0.5em \fi \nobreak
      \vrule height0.75em width0.5em depth0.25em\fi}

\providecommand{\DontPrintSemicolon}{\dontprintsemicolon}

\def\proof{{\noindent{\bf Proof of Algorithm: }}}
\def\revisedtext#1{\color{red} #1\normalcolor}

\newcommand{\R}{\mathbb{R}}

\newcommand{\N}{\mathbb{N}}

\newcommand{\be}{\begin{equation}}
\newcommand{\en}{\end{equation}}

\usepackage{tikz}
\usetikzlibrary{arrows,snakes,positioning,patterns}
\tikzstyle{vertex}=[circle,fill=white,draw=black,minimum size=18pt,inner sep=0pt]
\tikzstyle{cobrinha}=[decorate,decoration={snake,amplitude=.4mm,segment length=2mm,post length=1mm,pre length=2mm}] 
	
\title{Higher-order Reverse Automatic Differentiation with emphasis on the third-order.}
\author{R. Gower\footnote{School of Mathematics
and Maxwell Institute for Mathematical Sciences
The University of Edinburgh, e-mail: gowerrobert@gmail.com}  $\,\,\,\cdot\,$ A. Gower\footnote{School of Mathematics, Statistics and Applied Mathematics, National University of Ireland Galway}}

\begin{document}

\maketitle
\begin{abstract}
It is commonly assumed that calculating third order information is too expensive for most applications. But we show that the directional derivative of the Hessian ($D^3f(x)\cdot d$) can be calculated at a cost proportional to that of a state-of-the-art method for calculating the Hessian matrix. 
We do this by first presenting a simple procedure for designing high order reverse methods and applying it to deduce several methods including a reverse method that calculates $D^3f(x)\cdot d$. We have implemented this method taking into account symmetry and sparsity, and successfully calculated this derivative for functions with a million variables. 
These results indicate that the use of third order information in a general 
nonlinear solver, such as Halley-Chebyshev methods, could be a practical alternative to Newton's method. 
\end{abstract}
\section{Introduction}

Derivatives permeate mathematics and engineering right from the first steps of calculus, which together with the Taylor series expansion is a central tool in designing models and methods of modern mathematics. Despite this, successful methods for automatically calculating derivatives of $n$-dimensional functions is a relatively recent development.
 Perhaps most notably amongst recent methods is the advent of Automatic Differentiation (\emph{AD}), which has the remarkable achievement of the ``cheap gradient principle'', wherein the cost of evaluating the gradient is proportional to that of the underlying function~\cite{Griewank:2008}. This AD success is not only limited to the gradient, there also exists a number of efficient AD algorithms for calculating Jacobian~\cite{Gebremedhin:2005,Griewank2003} and Hessian matrices~\cite{Gower:2012,Gebremedhin:2009}, that can accommodate for large dimensional sparse instances. The same success has not been observed in calculating higher order derivatives. 

 The assumed cost in calculating high-order derivatives drives the design of methods, typically favouring the use of lower-order methods. 
In the optimization community it is generally assumed that calculating any third-order information is too costly, so the design of methods revolves around using first and second order information. 
 We will show that third-order information can be used at a cost  proportional to the cost of calculating the Hessian. This has an immediate application in third-order nonlinear optimization methods such as the Chebyshev-Halley Family~\cite{Gutierrez1997a}.
Furthermore, the need for higher order differentiation finds applications in calculating quadratures~\cite{Kariwala2012,Corliss1997}, bifurcations and periodic orbits~\cite{Guckenheimer2000}. In the fields of numerical integration and solution of PDE's, a lot of attention has been given to refining and adapting meshes to then use first and second-order approximations over these meshes. An alternative paradigm would be to use fixed coarse meshes and higher approximations. 
 With the capacity to efficiently calculate high-order derivatives, this approach could become competitive  
and lift the fundamental deterrent in higher-order methods.


Current methods for calculating derivatives of order three or higher in the AD community typically propagate univariate Taylor series~\cite{Griewank2000} or repeatedly apply the tangent and adjoint operations~\cite{Naumann2012TAo}. In these methods, each element of the
desired derivative is calculated separately. If AD has taught us anything it is that we should not treat elements of derivatives separately, for their computation can be highly interlaced. The cheap gradient principle illustrates this well, for calculating the elements of the gradient separately yields a time complexity of $n$ times that of simultaneously calculating all entries. This same principle should be carried over to higher order methods, that is, be wary of overlapping calculations in individual elements.
Another alternative for calculating high order derivatives is the use of forward
differentiation~\cite{Neidinger1992}. The drawback of forward propagation is that it calculates the derivatives of all intermediate functions, in relation to the independent variables, even when these do not contribute to the desired end result. For these reasons, we look at calculating high-order derivatives as a whole and focus on reverse AD methods.      

An efficient alternative to AD is that the end users hand code their derivatives. Though with the advent of evermore complicated models, this task is becoming increasingly error prone, difficult to write efficient code, and, let's face it, boring. This approach also rules out methods that use high order derivatives, for no one can expect the end user to code the total and directional derivatives of high order tensors.

The article flows as follows, first we develop algorithms that calculate derivatives in a more general setting, 
wherein our function is described as a sequence of compositions of maps, Section~\ref{sec:DerivativesOfCompositions}. We then use Griewank and Walther's~\cite{Griewank:2008} state-transformations in Section~\ref{sec:ImplementingStateTransform}, to translate a composition of maps into an AD setting and an efficient implementation. Numerical tests are presented in Section~\ref{sec:tests}, followed by our conclusions in Section~\ref{sec:Conclusion}.

\section{Derivatives of Sequences of Maps}
\label{sec:DerivativesOfCompositions}

In preparation for the AD setting, we first develop algorithms for calculating derivatives of functions
that can be broken into a composition of operators
\begin{equation} \label{eq:mapcomp}
F(x) = \Psi^{\ell} \circ \Psi^{\ell-1} \circ \cdots \circ \Psi^{1}(x).
\end{equation}
 for $\Psi^i$'s of varying dimension: $\Psi^1(x) \in C^2(\R^{n},\R^{m_1})$ and $\Psi^i(x)\in C^2(\R^{m_{i-1}},\R^{m_i})$, each  $m_i \in \N$ and for $i=2,\ldots, \ell$, so that $F: \R^{n} \rightarrow \R^{m_{\ell}}$. From this we define a functional $f(x) = y^T F(x),$ where $y\in \R^{m_{\ell}}$, and develop methods for calculating the \emph{gradient} $\nabla f(x) = y^T DF(x)$, the \emph{Hessian} $D^2 f(x) = y^T D^2F(x)$ and the \emph{Tensor} $D^3f(x) = y^T D^3F(x)$. 

  For a given $d \in \R^n$, we also develop methods for the directional derivative  $D F(x)\cdot d$, $D^2F(x)\cdot d$, the Hessian-vector product $D^2f(x)\cdot d = y^T D^2 F(x)\cdot d$ and the Tensor-vector product $D^3 f(x) \cdot d = y^T D^3 F(x)\cdot d$. Notation will be gradually introduced and clarified as is required, including the definition of the preceding directional derivatives.

\subsection{First-Order Derivatives}
Taking the derivative of $F$, equation~\eqref{eq:mapcomp}, and recursively applying the chain rule, we get
\begin{equation}
\label{eq:gradientmaps}
 y^T D F = y^T D\Psi^{\ell}D\Psi^{\ell-1} \cdots D\Psi^{1}.
\end{equation}
Note that $y^T D F$ is the transpose of the gradient $\nabla (y^T F ) $. For simplicity's sake, the argument of each function is omitted in (\ref{eq:gradientmaps}), but it should
be noted that $D\Psi^i$ is evaluated at the argument $(\Psi^{i-1}\circ\cdots\circ\Psi^{1})(x)$, for each
$i$ from $1$ to $\ell$. If each of these arguments has been recorded, the gradient of $y^TF(x)$ can be calculated with what's called a
\emph{reverse sweep} in Algorithm~(\ref{alg:gradientblock}). Reverse, for it transverses the maps from the last
$\Psi^{\ell}$ to the first $\Psi^{1}$, the opposite direction in which~(\ref{eq:mapcomp}) is evaluated.
The intermediate stages of the gradient calculation are accumulated in the vector $\overline{\mathrm{v}}$, its dimension changing from one iteration to the next. This will be a recurring fact in the matrices and vectors used to store the intermediate phases of the archetype algorithms presented in this article. 

\DontPrintSemicolon
\begin{algorithm}[H]
\caption{ Archetype Reverse Gradient.}
\label{alg:gradientblock}
\textbf{initialization}: $ \overline{\mathrm{v}} = y$\;
\For{$i = \ell, \ldots, 1$}{
$ \overline{\mathrm{v}}^T \leftarrow \overline{\mathrm{v}}^T D \Psi^i$
}
\KwOut{$ y^T D F(x) =  \overline{\mathrm{v}}^T$}
\end{algorithm}

For a given direction $d \in \R^n$, we define the directional derivative of $F(x)$ as
\begin{align} 
\label{eq:defdfv}
\frac{d}{dt} F(x+td) &=  D_i F(x+td) d_i  := DF(x+td)\cdot d, 
\end{align}
where we have omitted the summation symbol for $i$, and instead, use Einstein notation where a
repeated indexes implies summation over that index.
We use this notation throughout the article unless otherwise stated.
Again using the chain-rule and~(\ref{eq:mapcomp}), we find
\[DF(x)\cdot d = D\Psi^{\ell} D\Psi^{\ell-1} \cdots D\Psi^{1}\cdot d. \]
This can be efficiently calculated using a \emph{forward sweep} of the computational graph, detailed below.
\begin{algorithm}
\caption{Archetype 1st Order Directional Derivative.}
\label{alg:directionalblock}
\textbf{initialization}: $\dot{\mathrm{v}}^0 = d$\;
\For{$i = 1, \ldots, \ell$}{
$\dot{\mathrm{v}} \leftarrow D\Psi^i \dot{\mathrm{v}}$ 
}
\KwOut{$D F(x)\cdot d = \dot{\mathrm{v}}^{\ell}$}
\end{algorithm}

\subsection{Second-Order Derivatives}
Here we develop a reverse algorithm for calculating the Hessian $D^2 (y^T F(x))$. 
First we determine the Hessian for $F$ as a composition of two maps, then we use induction to design a method for when $F$ is a composition of $\ell$ maps. 

For $F(X) = \Psi^2 \circ \Psi^1(x) $ and $\ell=2$, we find the Hessian by differentiating in the $j$-th and $k$-th coordinate,
\begin{equation}\label{eq:hesscompein}
D_{jk}( y_i F_i )  =   (y_i D_{rs}\Psi_i^2 ) D_j\Psi_r^1  D_k\Psi_s^1 +  (y_i D_r\Psi^2_i )D_{jk}\Psi^1_r,
\end{equation}
where the arguments have been omitted. So the $(j,k)$ component of the Hessian $[ D^2(y^T F) ]_{jk}= D_{jk}(y^T F)$. The higher the order of the derivative, the more messy and unclear component notation becomes. A way around this issue is to use a tensor notation 
\begin{gather}
  y^T D^2 F\cdot ( v, w) :=  y_i D_{jk} F_i  v_j w_k,  \notag
\end{gather}
and 
\begin{gather}
   (y^T D^2 F\cdot  w) \cdot v := y^T D^2 F\cdot ( v, w) ,
\end{gather}

for any vectors $v,w \in \R^{n}$, and in general, 
\begin{gather}
   [y^T D^2 F \cdot ( \triangle, \square) ]_{t_2 \cdots t_q  s_2 \cdots s_p}  :=  y_i D_{t_1 s_1} F_i  \triangle_{t_1
t_2 \cdots t_q} \square_{s_1 s_2 \cdots s_p},
\end{gather}
and
\begin{gather}
   (y^T D^2 F\cdot \square ) \cdot \triangle  := y^T D^2 F\cdot ( \triangle, \square) 
\end{gather}
for any compatible $\triangle$ and $\square$ . To use a matrix notation for a composition of maps can be aesthetically
unpleasant. Using this tensor notation the Hessian of $y^T F$, see equation~\eqref{eq:hesscompein}, becomes
\begin{equation}
\label{eq:Hessian2map}
\boxed{
  y^T D^2 F = y^T D^2 \Psi^2 \cdot ( D\Psi^1,D\Psi^1 ) +y^T D \Psi^2 \cdot D^2\Psi^1 
}.
\end{equation}
We recursively use the identity~\eqref{eq:Hessian2map} to design an algorithm that calculates the Hessian of a function $y^TF(x)$ composed of $\ell$ maps, as defined in equation~(\ref{eq:mapcomp}).

\newpage

\begin{algorithm}
\caption{Archetype Reverse Hessian.}
\label{alg:edgepushingblock}
\textbf{initialization}: $\overline{\mathrm{v}} = y$, $W=0$\;
\For{$i = \ell, \ldots, 1$}{
$W\leftarrow W \cdot( D\Psi^i, D \Psi^i)$\;
$W\leftarrow W+\overline{\mathrm{v}}^T D^2\Psi^i$\;
$\overline{\mathrm{v}}^T \leftarrow \overline{\mathrm{v}}^T D\Psi^i$
}
\KwOut{$y^T D^2 F \leftarrow W, \,\, y^T DF \leftarrow \overline{\mathrm{v}}^T $}
\end{algorithm}
{\proof} We will use induction on the number of compositions $\ell$.
For $\ell=1$ the output is $W = y^T D^2\Psi^{1}$. Now we suppose the Algorithm is correct for $m-1$ map compositions, and use this assumption to show that for $\ell = m$ the output is $W = y^T D^2 F$. Let
\[y^T X = y^T \Psi^{m} \circ \cdots \circ \Psi^2,\]
so that $y^T F = y^T X \circ \Psi^1.$ Then at the end of the iteration $i=2$, by the chain rule, $\overline{\mathrm{v}}^{T} = y^T D X$ and, by induction, $W =y^T D^2 X$. This way, at termination, or after the iteration $i=1$, we get
\begin{align*}
W &= y^T D^2 X \cdot (D \Psi^1 ,D \Psi^1)  + y^T D X \cdot D^2 \Psi^1\\
  &=  y^T D^2(X \circ \Psi^{1}) \quad \quad \left[\mbox{Equation~\eqref{eq:Hessian2map}}\right]\\
  & = y^T D^2 F. \qed
\end{align*}

Now we take a small detour to show how to calculate Hessian-vector products in a similar manner. We do this because it is an important component of graph-coloring based algorithms for calculating the Hessian~\cite{Gebremedhin:2009} and its complexity is surprisingly the same as evaluating $y^T F$, the underlying functional~\cite{Christianson:1992}.
Thus, analogously, we calculate the directional derivative of the gradient $y^T DF(x)$, for $\ell =2$,
\begin{align} 
 y^TD_{jk}F d_k &=   y^T D_{rs}\Psi^2 D_j\Psi^1_r D_k\Psi^1_s d_k + y^T D_r\Psi^2  D_{jk} \Psi^1_r d_k, \notag \\
\end{align}
or simply,
 \begin{align}
 \label{eq:y2ndderein}
 \boxed{y^TD^2F \cdot d  =   y^T D^2\Psi^2 \cdot (D\Psi^1 ,D\Psi^1 \cdot d )+ y^T D\Psi^2 \cdot D^2 \Psi^1 \cdot d }, 
\end{align}
and use this recursively to calculate the directional derivative of $y^T DF(x)$ in Algorithm~\ref{alg:y2ndderein}. This algorithm was first described in~\cite{Christianson:1992}.


\begin{algorithm}
\caption{ Archetype Gradient Directional Derivative}
\label{alg:y2ndderein}
\textbf{initialization}: $\dot{\mathrm{v}}^0 = d, \overline{\mathrm{v}} =y \in \R^{m_{\ell}}, w=0\in
\R^{m_{\ell}}$\;
\For{$i = 1,\ldots, \ell$}{
$\dot{\mathrm{v}}^i \leftarrow D\Psi^i  \cdot \dot{\mathrm{v}}^{i-1}$ \;
}
\For{$i = \ell, \ldots, 1$}{
$w\leftarrow  w \cdot D \Psi^i $\;
$w\leftarrow w+\overline{\mathrm{v}}^T D^2 \Psi^i \cdot \dot{\mathrm{v}}^{i-1}$ \;
$\overline{\mathrm{v}}^T \leftarrow \overline{\mathrm{v}}^T D \Psi^i$
}
\KwOut{$y^T D^2 F(x)\cdot d \leftarrow w, \,\, y^T DF \leftarrow \overline{\mathrm{v}}^T$}
\end{algorithm}
{\proof}
Let  $y^TF$ be a composition of $\ell$ maps as in~(\ref{eq:mapcomp}) and
\[X^m = y^T\Psi^{\ell} \circ \cdots \circ \Psi^m,\]
so that  $X^{m-1} = X^m \circ \Psi^{m-1}$. 
 The first {\bf for} loop simply accumulates the directional derivative $D F \cdot d$. For the second {\bf for} loop, we use an induction hypothesis that at the end of the $i=m$ iteration $w = D^2 X^m \cdot \dot{\mathrm{v}}^{m-1}$. The first iteration, $i =\ell$, the output is $w = y^T  D^2 \Psi^{\ell} \cdot d = D^2 X^\ell \cdot \dot{\mathrm{v}}^{\ell-1}$. Now suppose our hypothesis is true for $i = m+1$, so that at the end of the $i=m+1$ iteration, by the induction hypothesis, 
 \[ 
 w = D^2 X^{m+1} \cdot \dot{\mathrm{v}}^{m} = D^2 X^{m+1} \cdot D\Psi^m \cdot \dot{\mathrm{v}}^{m-1},
 \] 
 and, by calculus,
\[\overline{\mathrm{v}}^T = y^T D \Psi^{\ell}  \cdots D \Psi^{m+1} = D X^{m+1}. \] 
Then for the next step, the $i=m$ iteration,
\begin{align*}
w &\leftarrow w \cdot D \Psi^{m} + \overline{\mathrm{v}}^T D^2\Psi^{m} \cdot \dot{\mathrm{v}}^{m-1}\\
  &= (D^2 X^{m+1} \cdot D\Psi^m \cdot \dot{\mathrm{v}}^{m-1}) \cdot D \Psi^m +DX^{m+1} \cdot D^2\Psi^{m} \cdot \dot{\mathrm{v}}^{m-1} \\
  &= D^2 X^{m+1} \cdot (D \Psi^m , D\Psi^m \cdot  \dot{\mathrm{v}}^{m-1})  +DX^{m+1} \cdot D^2\Psi^{m} \cdot \dot{\mathrm{v}}^{m-1} \\
   &= D^2( X^{m+1} \circ \Psi^{m}) \cdot \dot{\mathrm{v}}^{i-1} \quad \quad \left[\mbox{Equation~\eqref{eq:y2ndderein}}\right] \\
  &=  D^2 X^{m}\cdot  \dot{\mathrm{v}}^{m-1}.
\end{align*}

Thus by induction we have proved that at the end of the $i=1$ iteration,
\[w = D^2X^1\cdot \dot{\mathrm{v}}^0 = D^2X^{1}\cdot d = y^T D^2 F(x)\cdot d. \qed\]

\subsection{Third-Order Methods}


Now we move on to the directional derivative of $y^TD^2 F(x)$, that is, the derivative of $y^T D^2 F(x+td)$ in $t$, where $d \in \R^n$, to get
\begin{align}
 \frac{d}{dt}y^T D^2F(x+td) &= y_i \frac{d}{dt}D_{jk}F_i(x+td) \nonumber\\
&= y_i D_{jkm}F_i(x+td)d_m  \nonumber\\
&:= y^T D^3 F(x+td)\cdot d. \label{eq:tensorvdef}
\end{align}
Here our tensor notation really facilitates working with third-order derivatives. Using matrix notation would lead to confusing equations and possibly detter intuition.  The notation conventions from before carry over naturally to third-order derivatives, with 
\be
( y^T D^3 F \cdot ( \triangle,\square,\lozenge) )_{t_2 \ldots t_q s_2 \ldots s_p l_2 \ldots l_r}  := y_i D^3 F_{t_1 s_1 l_1} \triangle_{t_1  \ldots t_q} \square_{s_1\ldots s_p} \lozenge_{l_1\ldots l_r},
\en 
and
\be
\label{eq:ThirdOrderCommutation}
y^T D^3 F \cdot ( \triangle,\square,\lozenge) = (y^T D^3 F \cdot \lozenge) \cdot ( \triangle,\square) = ((y^T D^3 F \cdot \lozenge) \cdot  \square) \cdot \triangle,
\en 
for any compatible $\triangle$, $\square$ and $\lozenge$. We begin by calculating the directional derivative of a composition of two maps $F = \Psi^2 \circ \Psi^1$, 
\begin{align*}
\frac{d}{dt}& \left(y^T D^2 F (x+dt)\right) 
\\ 
  =& D \left ( y^T D^2 \Psi^2 \cdot(D\Psi^1,D\Psi^1) \right ) \cdot d + D \left ( ( y^T D\Psi^2 ) \cdot D^2\Psi^1\right ) \cdot d 
\\
  =& (y^T D^3\Psi^2 \cdot D\Psi^1 \cdot d)\cdot(D\Psi^1,D\Psi^1) +  ( y^T D^2\Psi^2)\cdot(D\Psi^1,D^2\Psi^1 \cdot d)  
\\
    & +( y^T D^2\Psi^2)\cdot(D^2\Psi^1 \cdot d,D\Psi^1) + (y^T D\Psi^2)\cdot D^3\Psi^1 \cdot d  
\\
    & + (y^T D^2\Psi^2 \cdot D\Psi^1 \cdot d )\cdot D^2\Psi^1,
\end{align*}
in conclusion, after some rearrangement,    
\begin{align}    
    y^T & \frac{d}{dt} D^2 F  (x+dt) 
 = y^T D^3\Psi^2 \cdot(D\Psi^1,D\Psi^1,D\Psi^1 \cdot d) + y^T D\Psi^2\cdot D^3\Psi^1 \cdot d 
 \notag \\
  &+  y^T D^2\Psi^2 \cdot\left ( (D\Psi^1,D^2\Psi^1\cdot d) + (D^2\Psi^1\cdot d,D\Psi^1) + (D^2\Psi^1,D\Psi^1 \cdot d)  \right )
    \label{eq:D3CompositionPsi2Psi1d}
\end{align}
As usual, we have omitted all arguments to the maps. The above applied recursively gives us the Reverse Hessian
Directional Derivative Algorithm~\ref{alg:RevHedir}, or \emph{RevHedir} for short. To prove the correctness of
\texttt{RevHedir}, we use induction based on
 $X^m = y^T \Psi^{\ell} \circ \cdots \circ \Psi^m$, working from $m=\ell$ backwards towards $m=1$ to calculate $y^T D^3
F(x)\cdot d$. 
\begin{algorithm}
\caption{Archetype Reverse Hessian Directional Derivative (\texttt{RevHedir} )}
\label{alg:RevHedir}
\textbf{initialization}: $\dot{\mathrm{v}}^1 = d, \overline{\mathrm{v}} =y, W =Td=0\in
\R^{m_{\ell}\times m_{\ell}}$\;
\For{$i = 1,\ldots, \ell$}{
$\dot{\mathrm{v}}^i \leftarrow D \Psi^i \cdot \dot{\mathrm{v}}^{i-1}$ \;
}
\For{$i = \ell, \ldots, 1$}{
$ Td \leftarrow Td \cdot ( D \Psi^i ,D \Psi^i ) $\;
$Td\leftarrow Td+ W \cdot \big( (D \Psi^i,D^2 \Psi^i\cdot \dot{\mathrm{v}}^{i-1})+ (D^2\Psi^i \cdot \dot{\mathrm{v}}^{i-1},D
\Psi^i ) \big) $\\
 $Td\leftarrow Td+ W \cdot ( D^2 \Psi^i,D \Psi^i \cdot \dot{\mathrm{v}}^{i-1})$\;
$Td\leftarrow Td+\overline{\mathrm{v}}^T  D^3\Psi^i \cdot \dot{\mathrm{v}}^{i-1} $\;
$W\leftarrow W \cdot(D \Psi^i,D \Psi^i) +\overline{\mathrm{v}}^T  D^2 \Psi^i$\;
$\overline{\mathrm{v}}^T \leftarrow \overline{\mathrm{v}}^T D \Psi^i$
}
\KwOut{$y^T D^3 F(x)\cdot d \leftarrow Td, \,\, y^T D^2 F \leftarrow W,\,\, y^T DF \leftarrow \overline{\mathrm{v}}^T$}
\end{algorithm}

\begin{proof} Our induction hypothesis is that at the end of the $i = m$ iteration $Td =   y^T D^3 X^{m} \cdot
\dot{\mathrm{v}}^{i-1} .$ After the first iteration $i=\ell$, paying attention to the initialization of the variables,
we have that $Td = \overline{\mathrm{v}}^T  D^3\Psi^{\ell} \cdot \dot{\mathrm{v}}^{\ell-1} = y^T D^3 X^{\ell} \cdot
\dot{\mathrm{v}}^{\ell-1}.$ Now suppose the hypothesis is true for iterations up to $m+1$, so that at the beginning of
the $i =m$ iteration $Td =y^T  D^3 X^{m+1} \cdot \dot{\mathrm{v}}^{m}.$ To prove the hypothesis we need the following
results: at the end of the $i=m$ iteration  
\begin{align}
 \overline{\mathrm{v}}^T & = y^T D X^{m} \quad \textrm{and} \quad W  = y^T D^2 X^{m}, \label{eq:tempW}
\end{align}
both are demonstrated in the proof of Algorithm~\ref{eq:y2ndderein}. Now we are equipt to examine $Td$ at the end of the
$i=m$ iteration,
\begin{align*}
 Td &\leftarrow Td \cdot ( D\Psi^m ,D\Psi^m ) + W \cdot \big( (D\Psi^m,D^2\Psi^m\cdot \dot{\mathrm{v}}^{m-1})+ (D^2\Psi^m\cdot
\dot{\mathrm{v}}^{m-1},D\Psi^m) \big)
 \\
&+  W \cdot (D^2\Psi^m,D\Psi^m\cdot \dot{\mathrm{v}}^{m-1}) + \overline{\mathrm{v}}^T  D^3\Psi^m \cdot
\dot{\mathrm{v}}^{m-1}, 
\end{align*} 
using the induction hypothesis followed by property~\eqref{eq:ThirdOrderCommutation} we get $Td \cdot ( D\Psi^m ,D\Psi^m
) = y^T  D^3 X^{m+1} \cdot \dot{\mathrm{v}}^{m} \cdot ( D\Psi^m ,D\Psi^m ) = y^T  D^3 X^{m+1} \cdot \left( D\Psi^m  ,
D\Psi^m ,\dot{\mathrm{v}}^{m}\right)$, and from the algorithm $\dot{\mathrm{v}}^{m}= D\Psi^m \cdot\dot{\mathrm{v}}^{m-1}
$. Then using equations~\eqref{eq:tempW} to substitute $W$ and $\overline{\mathrm{v}}^T$ we arrive at
\begin{align*}
Td= & y^T  D^3 X^{m+1} \cdot \left( D\Psi^m  , D\Psi^m ,D\Psi^m \cdot\dot{\mathrm{v}}^{m-1} \right) 
 \\
 &+ y^T D^2 X^m \cdot \big( (D\Psi^m,D^2\Psi^m\cdot \dot{\mathrm{v}}^{m-1})+ (D^2\Psi^m\cdot
\dot{\mathrm{v}}^{m-1},D\Psi^m) \big)
\\
&+  y^T D^2 X^m \cdot (D^2\Psi^m,D\Psi^m\cdot \dot{\mathrm{v}}^{m-1}) +  \left(y^T D X^m \right)D^3\Psi^m \cdot
\dot{\mathrm{v}}^{m-1}  
\\
= &  y^T D^3 X^m \cdot \dot{\mathrm{v}}^{m-1} \quad [\mbox{\bf Using equation~\ref{eq:D3CompositionPsi2Psi1d}}]. 
\end{align*}
Finally, after iteration $i=1$, we have
\[ Td = y^T D^3 X^1 \cdot \dot{\mathrm{v}}^0 = y^T D^3 F\cdot d. \qed\]
\end{proof}

As is to be expected, in the computation of the Tensor-vector product, only 2-dimensional tensor arithmetic, or matrix
arithmetic, is used, and it is not necessary to form a $3$-dimensional tensor. On the other hand,  calculating the
entire $y^T D^3 F$ Tensor does involve $3$-dimensional arithmetic. The final archetype algorithm we present is a reverse
method for calculating the entire third-order Tensor $y^T D^3 F(x)$. We want an expression for the derivative such that
 \be
 y^T \frac{d}{dt}D^2 F(x+td) =  y^T D^3 F(x+td)\cdot d
 \label{eq:D3toD3d}
 \en
 for any $d$. From equation~\eqref{eq:D3CompositionPsi2Psi1d}, we see that $d$ is contracted with the last coordinate in
every term except one. To account for this term, we need a \emph{switching tensor} $S$ such that
 \[
 y^T D^2\Psi^2 \cdot(D^2\Psi^1 \cdot d,D\Psi^1) = y^T D^2\Psi^2 \cdot(D^2\Psi^1  ,D\Psi^1) \cdot S \cdot d,
 \] 
in other words we define $S$ as
\be
\label{def:S}
S\cdot (v,w,z) =  (v,z,w) \quad \text{or} \quad S_{abcijk}v_i w_j z_k = v_a z_b w_c 
\en 
for any vectors $v,w$ and $z$. This implies that $S$'s components are $S_{abcijk} =
\delta_{ai}\delta_{cj}\delta_{bk}$, where $\delta_{nm} = 1$ if $n=m$ and $0$ otherwise.
 Then for $F = \Psi^2 \circ \Psi^1$ we use equation~\eqref{eq:D3CompositionPsi2Psi1d} to reach    
\begin{align}    
y^T & D^3 F \cdot d=   
    \left(  y^T D^3\Psi^2\cdot(D\Psi^1,D\Psi^1, D\Psi^1)  + y^T D\Psi^2 \cdot D^3\Psi^1  \notag \right.\\
     &+  \left. y^T D^2\Psi^2 \cdot \left ( (D\Psi^1,D^2\Psi^1 ) + (D^2\Psi^1 ,D\Psi^1)\cdot S +(D^2\Psi^1, D\Psi^1)
\right ) \right) \cdot d.
     \label{eq:D3CompositionPsi2Psi1}
\end{align}
The above is true for all vectors $d$, thus we can remove $d$ from both sides to arrive at our desired expression for
$y^T D^3 F.$
With this notation we have, as expected, $(y^T  D^3 F)_{ijk} = y^T D_{ijk}F $. We can now use this result to build a
recurrence for $D^3 X^m$, defined by $X^m = y^T \Psi^{\ell} \circ \cdots \circ \Psi^m$, working from $m=\ell$ backwards
towards $m=1$ to calculate $y^T D^3 F(x)\cdot d$. 
\begin{algorithm}
\caption{ Archetype Reverse Third Order Derivative}
\label{alg:Reverse3rd}
\textbf{initialization}: $\overline{\mathrm{v}} =y, W =0\in
\R^{m_{\ell}\times m_{\ell}}$, $T \in \R^{m_{\ell} \times m_{\ell} \times m_{\ell}} $\;
\For{$i = \ell, \ldots, 1$}{
$ T \leftarrow T \cdot ( D \Psi^i ,D \Psi^i,  D \Psi^i ) $\;
$T\leftarrow T+ W \cdot \big( (D \Psi^i,D^2 \Psi^i)+ (D^2\Psi^i ,D \Psi^i ) \big) $\\
 $T\leftarrow T+ W \cdot ( D^2 \Psi^i,D \Psi^i )\cdot S + \overline{\mathrm{v}}^T  D^3\Psi^i$\;
$W\leftarrow W \cdot(D \Psi^i,D \Psi^i) +\overline{\mathrm{v}}^T  D^2 \Psi^i$\;
$\overline{\mathrm{v}}^T \leftarrow \overline{\mathrm{v}}^T D \Psi^i$
}
\KwOut{$y^T D^3 F(x)\cdot d \leftarrow T,\,$ $y^T D^2 F \leftarrow W,\,$ $y^T DF \leftarrow \overline{\mathrm{v}}^T$}
\end{algorithm}

\proof the demonstration of this algorithm can be carried out in an analogous fashion to the proof of
Algorithm~\ref{alg:RevHedir}.

This notation, together with a closed expression for high-order derivatives of a composition of two maps,
see~\cite{Fraenkel2008a}, can be used to design algorithms of even higher-orders.
Though this would require the presentation of a rather cumbersome notation. What we can extract from this generic formula in~\cite{Fraenkel2008a}, is that the number of terms that need to be calculated 
grows combinatorially in the order of the derivative, thus posing a lasting computational challenge. 
 
\section{Implementing through State Transformations}
\label{sec:ImplementingStateTransform}
When coding a function, the user would not commonly write a composition of maps such as the form used in the previous
section, see equation~\eqref{eq:mapcomp}. Instead users implement functions in a number of different ways. Automatic
Differentiation (AD) packages standardize these hand written functions, through compiler tools and operator overloading,
into an evaluation that fits the format of Algorithm~\ref{alg:feval}. As an example, consider the function $f(x,y,z) =
xy\sin(z),$ and its evaluation for a given $(x,y,z)$ through the following list of commands
\begin{align*}
v_{-2} &= x\\
v_{-1} &= y\\
v_{0} &= z\\
v_{1} &= v_{-2}v_{-1}\\  
v_{2} &= \sin(v_{0}) \\ 
v_{3} &= v_2 v_1. 
\end{align*}
By naming the functions $\phi_1(v_{-2},v_{-1}) := v_{-2}v_{-1}$, $\phi_2(v_{0}) := \sin(v_0) $ and
 $ \phi_3(v_{2},v_{1}) :=v_2 v_1$, this evaluation fits the format in Algorithm~\ref{alg:feval}.

In general, each $\phi_i$ is an \emph{elemental function} such as addition, multiplication, $\sin(\cdot)$, $exp(\cdot)$,
etc,  which together with their derivatives are already coded in the AD package. In order, the algorithm first copies the
\emph{independent} variables $x_i$ into internal \emph{intermediate} variables $v_{i-n},$ for $i=1, \ldots, n$.
Following convention, we use negative indexes for elements that relate to independent variables. For consistency, we
will
  shift all indexes of vectors and matrices by $-n$ from here on, e.g., the components of $x \in \R^n$ are $x_{i-n}$
for $i =1 \ldots n.$

 The next step in Algorithm~\ref{alg:feval}, calculates the value $v_1$ that only depends on
the intermediate variables $v_{i-n},$ for $i =1, \ldots,n$. In turn, the value $v_2$ may now depend on $v_{i-n},$ for
$i=1, \ldots,n+1$, then $v_3$ may depend on $v_{i-n},$ for $i=1, \ldots,n+2$ and so on for all $\ell$ intermediate variables.
Each $v_i$ is calculated using
only one elemental function $\phi_i$.
There is a dependency amongst the intermediate variables, for
$\phi_i$ is evaluated at previously calculated intermediate variables.  
We say that $j$ is a predecessor of $i$ if $v_j$ is a necessary argument of $\phi_i$.
Let $P(i)$ be the set of predecessors of $i$ and
$v_{P(i)}$ the vector of predecessors, thus $\phi_i(v_{P(i)}) = v_i$ and necessarily $j <i$ for any $j \in P(i)$. 
Analogously, $S(i)$ is the set of successors of $i$.

\begin{algorithm}
\caption{ Function evaluation}
\label{alg:feval}
\KwIn{$v_{i-n} =x_i$, for $i =1, \ldots n$}
\For {$i = 1 \ldots \ell$ }{
   $v_i \leftarrow \phi_i(v_{P(i)})$}
	\KwOut{$f(x) \leftarrow v_{\ell}$}
\end{algorithm}
We can bridge this algorithmic description of a function with that of compositions of maps~(\ref{eq:mapcomp}) using
Griewank and Walther's \cite{Griewank:2008} state-transformations
\begin{align}
\Phi^i &\colon  \R^{n+\ell}\to\R^{n+\ell},\nonumber\\
& \parbox{25pt}{\centerline{$v$}} \mapsto  (v_{1-n},\ldots,v_{i-1},\phi_i(v_{P(i)}),v_{i+1},\ldots,
v_\ell)^T,\label{eq:statetrasnfPhi}
\end{align}
for $i =1, \ldots \ell$. In components,
\be
\label{eq:PhiCoordinates}
\Phi^i_r(v) = v_r(1 - \delta_{ri}) + \delta_{ri} \phi_i(v_{P(i)}),
\en
where here, and in the remainder of this article, we abandon Einstein's notation of repeated indexes, because having the limits of summation is useful for implementing.
With this, the function $f(x)$ defined by Algorithm~\ref{alg:feval} can be written as
\begin{equation}\label{eq:StateTransform}
 f(x) = e_{\ell+n}^T \Phi^{\ell} \circ \Phi^{\ell-1} \circ \cdots \circ \Phi^{1} \circ( P^{T} x),
\end{equation}
where 
$e_{\ell+n}$ is the $(\ell + n)$th canonical vector and $P$ is the immersion matrix $\left[I \,
\, 0\right]$ with $I \in \R^{n \times n}$ and $0 \in \R^{n\times (\ell -n) }.$
The Jacobian of the $i$th state transformation $\Phi^i$, in coordinates, is simply
\be
\label{eq:DPhiCoordinates} 
 D_j \Phi^i_r(v)= \delta_{rj}(1 - \delta_{ri}) + \delta_{ri}\frac{\partial \phi_i}{\partial v_j} (v_{P(i)}).
\en
%
   
With the state-transforms and the structure of their derivatives, we look again at a few of the archetype algorithms in
Section~\ref{sec:DerivativesOfCompositions} and build a corresponding implementable version. Our final goal
is to implement the \texttt{RevHedir} algorithm~\ref{alg:RevHedir}, for which we need the implementation of the reverse gradient and Hessian algorithms.

\subsection{First-Order}
\label{sec:ImplementFirstOrder}

To design an algorithm to calculate the gradient of $f(x)$, given in equation~\eqref{eq:StateTransform}, we turn to the
Archetype Reverse Gradient Algorithm~\ref{alg:gradientblock} and identify\footnote{Especifically $P^T$ would be
$\Psi^1$ and $\Phi^{i}$ would be $\Psi^{i+1}$.} the $\Phi^i$'s in place of the $\Psi^i$'s. Using~\eqref{eq:DPhiCoordinates}
 we find that $\overline{\mathrm{v}}^T \leftarrow \overline{\mathrm{v}}^T D{\Phi^i}$ becomes
 \be
 \bar{v}_j \leftarrow \bar{v}_j (1 - \delta_{ij}) + \bar{v}_i \frac{\partial \phi_i}{\partial v_j}(v_{P(i)}) \quad
\forall j \in \{1-n, \ldots, \ell\}
 \en
 where $\bar{v}_i$ is the $i$-th component of $\overline{\mathrm{v}}$, also known as the $i$-th \emph{adjoint} in the AD
literature.  Note that if $j\neq i$ in the above,  then the above step will only alter $\bar{v}_j$ if $j \in P(i).$
Otherwise if $j=i$,  
 then this update is equivalent to setting $\bar{v}_i =0.$ We can disregard this update, as $\bar{v}_i$ will not
be used in subsequent iterations. This is because $i \not \in P(m)$, for $m \leq i$.  With these considerations, we
arrive at the Algorithm~\ref{alg:gradientevaluationcomponent}, the component-wise version of
Algorithm~\ref{alg:gradientblock}. Note how we have used
the abbreviated operation $a+=b$ to mean $a\leftarrow a+b.$ Furthermore, the last step  $\overline{\mathrm{v}}^T \leftarrow
\overline{\mathrm{v}}^T P^T$ selects the adjoints corresponding to independent variables.
 
An abuse of notation that we will employ throughout, is that
whenever we refer to $\bar{v}_i$ in the body of the text, we are referring to the value of $\bar{v}_i$ after iteration
$i$ of the Reverse Gradient algorithm has finished. 

\begin{algorithm}
 \caption{ Reverse Gradient.}
 \label{alg:gradientevaluationcomponent}
\textbf{initialization}: $\overline{\mathrm{v}} = e_1 \in \R^{\ell +n}$\;
\For{$i = \ell, \ldots, 1$}{
\lFor{$j \in P(i)$}{
$\bar{v}_j += \bar{v}_i\partial \phi_i(v_{P(i)})/\partial v_j$}}
\KwOut{$\nabla f \leftarrow \overline{\mathrm{v}}^T P^T = (\bar{v}_{1-n},\ldots,\bar{v}_0)^T$}
\end{algorithm}

Similarly, by using~\eqref{eq:DPhiCoordinates} again, 
 each iteration $i$  of
the Archetype 1st Order Directional Derivative
Algorithm~\ref{alg:directionalblock}, can be reduced to a coordinate form 
\[
\dot{v}_r \leftarrow (1 - \delta_{ri}) \dot{v}_r + \delta_{ri} \sum_{j \in P(i)}\dot{v}_j \frac{\partial
\phi_i}{\partial v_j}(v_{P(i)}),
\]
 where $\dot{v}_j$ is the $j$-th component of $\dot{\mathrm{v}}$. If $r \neq i$ in the above, then $\dot{v}_r$ remains unchanged, while if $r=i$ then we have
\be
\label{eq:DPhidCoordinates} \dot{v}_i \leftarrow \sum_{j \in P(i)}\dot{v}_j \frac{\partial \phi_i}{\partial v_j}(v_{P(i)}).
\en
We implement this update by sweeping through the successors of each intermediate variable and incrementing a single term
to the sum on the right-hand side of~\eqref{eq:DPhidCoordinates}, see Algorithm~\ref{alg:directionalevaluationcomponent}. It is crucial to observe that the $i$-th component of
$\dot{\mathrm{v}}$ will remain unaltered after the $i$-th iteration.  
 
 Again, when we refer to $\dot{v}_i$ in the body of the text from this point on, we are referring to the value of
$\dot{v}_i$ after iteration $i$ has finished in Algorithm~\ref{alg:directionalevaluationcomponent}.

\begin{algorithm}
 \caption{1st Order Directional Derivative.}
 \label{alg:directionalevaluationcomponent}
\textbf{initialization}: $\dot{v}=P^Td \in \R^{\ell +n}$\;
\For{$j = 1, \ldots, \ell$}{
\lFor{$i \in S(j)$}{
$\dot{v}_i += \dot{v}_j\partial \phi_i(v_{P(i)})/\partial v_j$}}
\KwOut{$DF \cdot d = (\dot{v}_{1-n},\ldots,\dot{v}_0)^T$}
\end{algorithm}

\subsection{Second-Order}
Just by substituting $\Psi^i$s for $\Phi^i$s in the Archetype Reverse Hessian, Algorithm~\ref{alg:edgepushingblock}, we
can quickly reach a very efficient component-wise algorithm for calculating the Hessian of $f(x)$, given in
equation~\eqref{eq:StateTransform}. This component-wise  algorithm is also known as \texttt{edge\_pushing}, and has
already been detailed in Gower and Mello~\cite{Gower:2012}. Here we use a different notation which leads to a more
concise presentation. Furthermore, the results below form part of the calculations needed for third order methods. 

There are two steps of Algorithm~\ref{alg:edgepushingblock} we must investigate, for we already know how to update $\overline{\mathrm{v}}$ from the above section. For these two steps, we need to substitute
\be \label{eq:D2PhiCoordinates}
D_{jk}\Phi^i_r(v) = \frac {\partial^2 \Phi^i_r}{\partial v_j \partial v_k }(v) = \delta_{ri}\frac{\partial^2
\phi_i}{\partial v_j \partial v_k}(v_{P(i)}),
\en
and $D\Phi^i$, equation~\eqref{eq:DPhiCoordinates}, in $W \leftarrow W \cdot (D\Phi^{i},D\Phi^{i})+  \overline{\mathrm{v}}^T
D^2\Phi^{i}$, resulting in
\begin{align}
W_{jk} \leftarrow & \sum_{s,t=1-n}^{\ell}\frac {\partial \Phi^{i}_s}{\partial v_j}  W_{st} \frac {\partial \Phi^{i}_t}{\partial v_k}  + \sum_{s=1-n}^{\ell} \bar v_{s} \frac{\partial^2  \Phi^{i}_s}{\partial v_j \partial v_k} 
\notag \\
    =&(1 - \delta_{ji})  W_{jk} (1 - \delta_{ki})+ \frac{\partial \phi_{{i}}}{\partial v_j} W_{ii} \frac{\partial
\phi_{i}}{\partial v_k} \notag
 \\
   & +  \frac{\partial \phi_{i}}{\partial v_j} W_{ik} (1 - \delta_{ki})  + (1 - \delta_{ji}) W_{ji} \frac{\partial
\phi_{i}}{\partial v_k} \label{Up:WComponentEdgePush} 
   \\ 
  & +  \bar v_{i} \frac{\partial^2 \phi_{i}}{\partial v_j \partial v_k}.
   \label{Up:WComponentCreate} 
\end{align}
 Before translating these updates into an algorithm, we need a crucial result: at the beginning of iteration $i-1$, the element $W_{jk}$ is zero if $j \geq i$ for all $k$. We show this by using induction on
the iterations of Algorithm~\ref{alg:edgepushingblock}. Note that $W$ is initially set to zero, so the first step from~\eqref{Up:WComponentEdgePush} and~\eqref{Up:WComponentCreate} reduce to
\[ 
W_{jk} \leftarrow \bar v_{\ell} \frac{\partial^2 \phi_{\ell}}{\partial v_j \partial v_k}, 
\]
which is zero for $j = \ell$ because $\ell \notin P(\ell)$. Now we assume the induction hypothesis holds at the beginning of the iteration $i$, so that $W_{jk} = 0$ for $j \geq i+1$. So letting $j \geq i+1$ and executing the iteration $i$ we get from the updates~\eqref{Up:WComponentEdgePush} and~\eqref{Up:WComponentCreate} 
\[ 
W_{jk} \leftarrow W_{jk}+ W_{ji} \frac{\partial \phi_{i}}{\partial v_k}, 
\]
because $j \notin P(i)$. Together with our hypothesis $W_{jk} = 0$ and $W_{ji}=0$, we see that $W_{jk}$ remains zero. While if $j=i$, then~\eqref{Up:WComponentEdgePush} and~\eqref{Up:WComponentCreate} sets $W_{jk}  \leftarrow 0$ because $i \not\in P(i)$. 
Hence at the beginning of iteration $i-1$ we have that $W_{jk}=0$ for $j\geq i$ and this completes the induction.

 Furthermore, $W$ is symmetric at the beginning of iteration $i$ 
because it is initialized to $W =0$ and each iteration preserves symmetry. Consequentially, the only nonzero
components $W_{jk}$ appear when both $j,k \leq i.$  We make use of this symmetry to avoid unnecessary calculations on symmetric counterparts.  Let $W_{\{jk\}}$  denote both $W_{jk}$ and $W_{kj}.$ 
To accommodate for this symmetric representation, we perform~\eqref{Up:WComponentCreate} once for each pair $\{j,k\}$,
as to opposed for every coordinate pair, see the \texttt{Creating} step in
Algorithm~\ref{alg:edgepushingcomponentwise}. 

The calculations in~\eqref{Up:WComponentEdgePush} are done by sweeping through the nonzero elements of $W$ 
and then updating their contribution to the overall calculation.

Thus if $W_{\{ii\}} \neq 0$, looking to~\eqref{Up:WComponentEdgePush}, this triggers the following increment
\[W_{\{jk\}}+=  \frac{\partial \phi_{i}}{\partial v_j} W_{\{ii\}} \frac{\partial \phi_{i}}{\partial v_k}. \]
 Similarly to \texttt{Creating} step, the above should only be carried
out for every pair $\{j,k\}.$ 
While each nonzero off diagonal term $W_{ik}$ and $W_{ki}$, for $k <i$, according to~\eqref{Up:WComponentEdgePush}, has
the effect of
\begin{align}
W_{jk} &+=  \frac{\partial \phi_{i}}{\partial v_j} W_{ik}, \label{Up:pushingside1}\\
W_{kj} & +=  \frac{\partial \phi_{i}}{\partial v_j} W_{ki}. \label{Up:pushingside2}
\end{align} 
It is redundant to update both symmetric elements, 
so we substitute both for just
\[W_{\{jk\}} += \frac{\partial \phi_{i}}{\partial v_j} W_{\{ik\}}. \] 
Though we must take care when $j=k$, for according to~\eqref{Up:pushingside1} and~\eqref{Up:pushingside2}, the two symmetric updates ``double up'' on the diagonal
\begin{align}
W_{\{jj\}} &+=  
 2\frac{\partial \phi_{i}}{\partial v_j}W_{\{ij\}}.
\label{Up:WComponentDistilledSym}
 \end{align}
The operation~\eqref{Up:WComponentEdgePush} has been implemented with these above considerations in the \texttt{Pushing}
step in Algorithm~\ref{alg:edgepushingcomponentwise}. The names of the steps \texttt{Creating} and \texttt{Pushing} are
elusive to a graph interpretation~\cite{Gower:2012}.

\begin{algorithm}
 \caption{component-wise form of \texttt{edge\_pushing}.}
 \label{alg:edgepushingcomponentwise}
\KwIn{Function evaluation~\ref{alg:feval}, $x \in \R^n.$}
\textbf{initialization}: $\bar{v}=e_{\ell+n} \in \R^{\ell +n}$, $W = 0 \in \R^{(\ell+n)\times (\ell +n)}$\;
\For{$i = \ell, \ldots, 1$}{ 
\texttt{Pushing}\;
\ForEach{ $k \leq i$ such that $W_{\{ki\}}\neq 0$}
{
\eIf{$k< i$}
      {\ForEach{$j \in P(i)$}
			 {
			    \eIf{$j=k$}{$W_{\{jj\}} += 2D_j\phi_i W_{\{ji\}}$}
					{$W_{\{jk\}} +=D_j \phi_i W_{\{ki\}}$} 
			 }
  } (~$k=i$) {
  \ForEach{unordered pair $\{j,p\} \subset P(i)$}{
   $W_{\{jp\}} +=D_p \phi_i D_j \phi_i W_{\{ii\}}$}
	   }
  }
\texttt{Creating}\;
    \ForEach{unordered pair $\{j,p\} \subset P(i)$}{
        $W_{\{jp\}}+=\bar{v}_i D_{pj}\phi_i $
		}
\texttt{Adjoint}\;
  \ForEach{$j \in P(i)$}{$\bar{v}_j+=\bar{v}_i D_j \phi_i $}
}                              
	\KwOut{$D^ 2f=\left(W_{jk}\right)_{1-n \leq j,k\leq 0}$}

\end{algorithm}

\subsection{Third-Order}

The final algorithm that we translate to implementation is the Hessian directional derivative, the \texttt{RevHedir}
Algorithm~\ref{alg:RevHedir}. This implementation has an immediate application in the Halley-Chebyshev class of
third-order optimization methods, for
at each step of these algorithms, such a directional derivative is required. For this reason we pay special attention to
its implementation.

Identifying each $\Psi^i$ with $\Phi^i$, we address each of the five operations on the matrix $Td$ in
Algorithm~\ref{alg:RevHedir} separately, pointing out
how each one preserves the symmetry of $Td$ and how to perform the component-wise calculations.

First, given that $Td$ is symmetric, the \emph{2D pushing} update
\be
\label{eqn:RevHedirPushing}
Td\leftarrow  Td \cdot \left(D\Phi^i,D\Phi^i\right), 
\en
is exactly as detailed in~\eqref{Up:WComponentEdgePush} and the surrounding comments. While \emph{3D creating}
\[Td\leftarrow Td + \overline{\mathrm{v}}^T D^3\Phi^i \cdot \dot{\mathrm{v}}^{i-1}, \]
can be written in coordinate form as
\begin{align}
Td_{jk} &\leftarrow  Td_{jk}+ \sum_{r,p=1-n}^{\ell} \overline{v}_r D_{jkp}\Phi^i_r
\dot{\mathrm{v}}_p^{i-1} \nonumber\\
&=  Td_{jk}+ \sum_{p \in P(i)}\overline{v}_i \frac{\partial^3 \phi_i}{\partial v_j \partial v_k \partial v_p}
\dot{v}_p, \label{eq:3dcreating}
\end{align}
where $\dot{v}_p$ is the value  given to $\dot{v}_p$ after iteration $p$ in Algorithm~\ref{alg:directionalevaluationcomponent}. Note that $\dot{\mathrm{v}}_p^{i-1} = \dot{v}_p$ for $p \in
P(i)$, because $p \leq i-1$, so on the iteration $i-1$ of Algorithm~\ref{alg:directionalevaluationcomponent} the calculation of $\dot{v}_p$ will already have been finalized.
  Another trick we employ is that, since the above calculation is performed on iteration $i$, we know that $\bar{v}_i$
has already been calculated. These substitutions involving $\bar{v}_i$s and $\dot{v}_i$s will be carried out in the rest of the text with little or no comment. The update~(\ref{eq:3dcreating}) also preserves the symmetry of $Td$.

To examine the update,
\begin{equation} \label{eq:block2dconnect1}Td\leftarrow Td+  W \cdot\left(
D\Phi^i, D^2\Phi^i\cdot \dot{\mathrm{v}}^{i-1} \right),\end{equation}
we use~\eqref{eq:DPhiCoordinates} and~\eqref{eq:D2PhiCoordinates} to obtain the coordinate form 
\begin{align}
Td_{jk} & \leftarrow Td_{jk} +
\sum_{r,s=1-n}^{\ell}W_{rs}\left( \delta_{rj}(1-\delta_{ri})+ \delta_{ri}\frac{\partial \phi_i}{\partial v_j} \right)
\delta_{si}\frac{\partial^2 \phi_i}{\partial v_k \partial v_p} \dot{v}_p \nonumber \\
&=   Td_{jk}+ W_{ji}(1-\delta_{ji})\frac{\partial^2 \phi_i}{\partial v_k \partial v_p}  \dot{v}_p+ W_{ii}\frac{\partial \phi_i}{\partial v_j} \frac{\partial^2 \phi_i}{\partial v_k \partial v_p}  \dot{v}_p.
\label{eq:2dconnect1}
\end{align}
Upon inspection, the update
\[Td\leftarrow Td+ W \cdot \left( D^2\Phi^i\cdot \dot{\mathrm{v}}^{i-1}, D\Phi^i \right) \]
is the transpose of~(\ref{eq:2dconnect1}) due to the symmetry of $W$. So it can be written in coordinate form
as
\begin{align}
Td_{jk} 
&\leftarrow   Td_{jk}+ W_{ik}(1-\delta_{ki})\frac{\partial^2 \phi_i}{\partial v_j \partial v_p}  \dot{v}_p+ W_{ii} \frac{\partial \phi_i}{\partial v_k} 
\frac{\partial^2 \phi_i}{\partial v_j \partial v_p}  \dot{v}_p.
\label{eq:2dconnect2}
\end{align}
Thus update~(\ref{eq:2dconnect2}) together with~(\ref{eq:2dconnect1}) is equivalent to summing a symmetric matrix to $Td$, so the symmetry of $Td$ is still preserved.

Last we translate
\begin{equation} \label{eq:block2dconnect3}Td \leftarrow Td+ W \cdot \left( D^2\Phi^i, D\Phi^i \cdot \dot{\mathrm{v}}^{i-1}
\right), \end{equation}
to its coordinate form
\begin{align}
 Td_{jk} & \leftarrow Td_{jk}+\sum_{r,s=1-n}^{\ell} W_{rs}\delta_{ri} D_{jk}\Phi^i_r \left( \delta_{sp}(1-\delta_{si})+
\delta_{si}\frac{\partial \phi_i}{ \partial v_p}\right)
 \dot{v}_p \nonumber \\
 &= Td_{jk}+ W_{ip}\frac{\partial^2 \phi_i}{\partial v_j \partial v_k}  (1-\delta_{pi})\dot{v}_p
 + W_{ii}\frac{\partial^2 \phi_i}{\partial v_j \partial v_k} D_{p} \phi_i
 \dot{v}_p. \label{eq:2dconnect3}
\end{align}
 
No change is affected by interchanging the indices $j$ and $k$ on the right-hand side of~(\ref{eq:2dconnect3}), so once again $Td$ remains symmetric. For convenience of computing, we group
updates~(\ref{eq:2dconnect1}),~(\ref{eq:2dconnect2})
and~(\ref{eq:2dconnect3}) into a set of updates called \texttt{2D Connecting}. The name indicating that
these updates ``connect'' objects that contain second order derivative information. 
   
More then just symmetric, through closer inspection of these operations, we see that the sparsity structure of $Td$ is contained in that of $W$. This remains true even after execution, at which point $Td= D^3f(x)\cdot d$ and $W = D^2f(x)$ where, for each $j,k,p \in \{1-n, \ldots, 0\}$, we have
\begin{equation*}
D_{jk}f(x) = 0 \implies
D_{jkp}f(x)d_p = 0.
\end{equation*}
This fact should be explored when implementing the method, in that, the data structure of $Td$ should imitate that of $W$.

\subsubsection{Implementing Third-Order Directional Derivative}

The matrices $Td$ and $W$ are symmetric, and based on the assumption that they will be sparse, we will
represent them using a symmetric sparse data structure. Thus we now identify each pair $(W_{jk},W_{kj})$ and
$(Td_{jk},Td_{kj})$ with the element $W_{\{jk\}}$ and $Td_{\{jk\}}$, respectively.
 Much like was done with \texttt{edge\_pushing},
Algorithm~\ref{alg:edgepushingcomponentwise}, we will organize the computations by sweeping through all nonzero elements of $Td_{\{ik\}}$ and $W_{\{ik\}}$ and then updating their contribution to the overall calculation.

We must take care when updating our symmetric representation of $Td$, both for the 2D pushing update~\eqref{eqn:RevHedirPushing} and for the redundant symmetric counterparts~\eqref{eq:2dconnect1} and~\eqref{eq:2dconnect2} which ``double-up'' on the diagonal, much like in the \texttt{Pushing} operations of Algorithm~\ref{alg:edgepushingcomponentwise}.


Each operation ~(\ref{eq:2dconnect1}),~(\ref{eq:2dconnect2})
and~(\ref{eq:2dconnect3}) depends on a diagonal element $W_{\{ii\}}$ and an off-diagonal element $W_{\{ik\}}$ of $W$, for $k \neq i$. Grouping together all terms that involve $W_{\{ii\}}$ we get the resulting update
\begin{align}
 Td_{\{jk\}}&+=   W_{\{ii\}}  \sum_{p \in P(i)}\dot{v}_p\left( \frac{\partial \phi_i}{\partial v_j}
\frac{\partial^2 \phi_i}{\partial v_k \partial v_p}  + \frac{\partial \phi_i}{\partial v_k } \frac{\partial^2
\phi_i}{\partial v_j \partial v_p}
+ \frac{\partial \phi_i}{ \partial v_p} \frac{\partial^2 \phi_i}{\partial v_j \partial
v_k}\right).\label{eq:2dconnect12} 
\end{align}
By appropriately renaming the indices in~(\ref{eq:2dconnect1}),~(\ref{eq:2dconnect2})
and~(\ref{eq:2dconnect3}), each nonzero off diagonal elements $W_{\{ik\}}$ gives the updates
~(\ref{eq:2dconnect1wik}),~(\ref{eq:2dconnect2wik})
and~(\ref{eq:2dconnect3wik}), respectively.
\begin{align}
 Td_{jk}&+=   \sum_{p \in P(i)} \dot{v}_p \frac{\partial^2 \phi_i}{\partial v_j \partial v_p} W_{ik},  \quad \forall j
\in P(i)    \label{eq:2dconnect1wik} \\
 Td_{kj}&+=  \sum_{p \in P(i)} \dot{v}_p \frac{\partial^2 \phi_i}{\partial v_j \partial v_p} W_{ki},  \quad \forall j
\in P(i)   \label{eq:2dconnect2wik} \\
 Td_{jp}&+=  \sum_{p \in P(i)} \dot{v}_k \frac{\partial^2 \phi_i}{\partial v_j \partial v_p} W_{ik}, \quad \forall j \in
P(i)   \label{eq:2dconnect3wik}
\end{align}  
Note that~(\ref{eq:2dconnect1wik}) and~(\ref{eq:2dconnect2wik}) are symmetric updates, and when $j=k$ these two operations ``double-up'' resulting in the update 
\[Td_{jj} += 2 \sum_{p \in P(i)} \dot{v}_p \frac{\partial^2 \phi_i}{\partial v_j \partial v_p} W_{ij}.  \]
Passing to our symmetric notation, we dispense with~(\ref{eq:2dconnect2wik}) and account for this doubling effect in Algorithm~\ref{alg:Tdcomponent-wise}.
Finally we can eliminate redundant symmetric calculations performed in~(\ref{eq:2dconnect3wik}) by only performing this
operation for each pair $\{j,p\}$.
 All these considerations relating to \texttt{2D connecting} have been factored into our implementation of the
\texttt{RevHedir} Algorithm~\ref{alg:Tdcomponent-wise}.

 Performing \texttt{3D Creating}~\eqref{eq:3dcreating} using this symmetric representation is
simply a matter of not repeating the obvious symmetric counterpart, but instead, performing these operations on
$Td_{\{jk\}}$ once for each appropriate pair $\{j,k\},$ see \texttt{3D Creating} in to
Algorithm~\ref{alg:Tdcomponent-wise}.


\begin{algorithm}
 \caption{component-wise form of \texttt{RevHedir}.}
 \label{alg:Tdcomponent-wise}

\KwIn{Function evaluation~\ref{alg:feval}, $x \in \R^n.$}
\textbf{Initialization}: $\bar{v}_{1-n} = \cdots=\bar{v}_{\ell-1}=0$, $\bar{v}_\ell = 1$, $W_{jk}=0$, $Td_{\{jk\}}=0$,
$j<k \in \{1-n, \ldots, \ell\}$ \;
\textbf{Calculate} first order directional derivative $\dot{v}$ using
Algorithm~(\ref{alg:directionalevaluationcomponent})\; 
\For{$i = \ell, \ldots, 1$}{
\texttt{2D Pushing} of $Td$, see \texttt{Pushing} in Algorithm~\ref{alg:edgepushingcomponentwise}\;
\texttt{2D Connecting}\;
   \ForEach{$p \in P(i)$, $\{j,k\} \subset P(i)$}
             {
             $Td_{\{jk\}} += W_{\{ii\}}\dot{v}_p \left(D_j\phi_i D_{kp}\phi_i + D_k \phi_i D_{jp}\phi_i+
D_p\phi_iD_{jk}\phi_i \right)$ \;
							
        }
   \ForEach{$k <i, W_{\{ik\}} \neq 0$}{
			\ForEach{$(j,p) \in P(i)^2$}{
			\If{$j=k$}{
			$Td_{\{kk\}} += 2W_{\{ik\}}\dot{v}_pD_{jp}\phi_i$\;
			}
			\If{$j \neq k$}{
			$Td_{\{jk\}} += W_{\{ik\}}\dot{v}_p D_{jp}\phi_i$\;
			}
			\If{$j \geq p$}{
			  $Td_{\{jp\}} += W_{\{ik\}}\dot{v}_k D_{jp}\phi_i $ \;
			 }
		}
}
\texttt{3D Creating}\;
   \ForEach{$p \in P(i)$, $\{j,k\} \subset P(i)$}
             {
             $Td_{\{jk\}} += \overline{v}_i D_{jkp}\phi_i \dot{v}_p $
        }
  \texttt{Pushing and creating} applied to $W$, see Algorithm~\ref{alg:edgepushingcomponentwise}\;
	\texttt{Adjoint Iteration} applied to $\bar{v}$, see Algorithm~\ref{alg:gradientevaluationcomponent}\;
}
\KwOut{$(D^3f(x)\cdot d)_{jk} = Td_{\{jk\}}, D^2f(x)_{jk} = W_{\{jk\}}$ \\ $\mbox{ for each } j\leq k \in \{1-n, \ldots,
0\}.$}

\end{algorithm}

\section{Numerical experiment} 
\label{sec:tests}
We have implemented the \texttt{RevHedir} Algorithm~\ref{alg:Tdcomponent-wise} as an additional driver of ADOL-C, a well
established automatic differentiation library coded in C and C++~\cite{Griewank:1996}. We
used version ADOL-C-2.4.0, the most recent available~\footnote{As checked May 28th, 2013}. The tests where carried out on a personal laptop with 1.70GHz dual core processors Intel Core i5-3317U, 4GB of RAM, with the Ubuntu 13.0 operating system.

For those interested in replicating our implementation, we used a sparse undirected weighted graph data structure to represent the matrices~$W$ and $Td$. The data structure is an array of weighted neighbourhood sets, one for each node,
where each neighbourhood set is a dynamic array that resizes when needed. Each neighbourhood set is maintained in order
and the method used to insert or increment the weight of an edge is built around a binary search. 

We have hand-picked sixteen problems from the CUTE collection \cite{Bongartz:1995}, \textsf{augmlagn} 
from \cite{Hock:1980}, \textsf{toiqmerg} (Toint Quadratic Merging problem) and
\textsf{chainros\_trigexp} (Chained Rosenbrook function with Trigonometric and exponential constraints) from
\cite{LadislavLuksanJanVlcek2003} for the experiments. We have also created a function 
\[\mbox{heavey\_band($x,band$)} = \sum_{i=1}^{n-band} \sin\left(\sum_{j=1}^{band} x_{i+j} \right). \]
For our experiments, we tested \texttt{heavey\_band($x,20$)}.  
The problems were selected based on the sparsity pattern of $D^3f(x).d$, 
dimension scalability and sparsity. Our goal was to cover a variety of patterns, to
easily change the dimension of the function and work with sparse matrices. 

In Table~\ref{tab:tests}, the ``Pattern'' column indicates the type of sparsity pattern: bandwidth\footnote{The bandwidth of matrix $M=(m_{ij})$ is the maximum value of $2|i-j|+1$ such that $m_{ij}\neq 0$.} of value $x$ (B $x$), arrow, frame, number of diagonals (D $x$),  or irregular pattern. The ``nnz/n''
column gives the number of nonzeros in $D^3f(x).d$ over the dimension $n$, which serves as a of measure density. For each problem, we applied \texttt{RevHedir} and \texttt{edge\_pushing} Algorithm~\ref{alg:Tdcomponent-wise} and~\ref{alg:edgepushingcomponentwise} to the objective function  $f: \R^n \rightarrow \R$, with $x_i = i$ and $d_i=1$, for $i =1,\ldots, n$, and give the runtime of each method for dimension $n=10^6$ in Table~\ref{tab:tests}. Note that all of these matrices are very sparse, partly due to the ``thinning out'' caused by the high order differentiation. This probably contributed to the
relatively low runtime, for in these tests, the run-times have a $0.75$ correlation with the density measure ``nnz/n''. This leads us to believe that the
actual pattern is not a decisive factor in runtime.

\begin{table}									
\centering									
\footnotesize									
\begin{tabular}{|l|c|rrr|}					\hline				
name	&	Pattern	&	nnz/n	&	\texttt{edge\_pushing}	&	\texttt{RevHedir}	\\ \hline \hline
cosine	&	B 3	&	3.0000	&	2.89	&	5.25	\\
bc4	&	B 3	&	3.0000	&	3.93	&	7.87	\\
cragglevy	&	B 3	&	2.9981	&	5.41	&	10.6	\\
chainwood	&	B 3	&	1.4999	&	4.04	&	7.22	\\
morebv	&	B 3	&	3.0000	&	4.57	&	9.44	\\
scon1dls	&	B 3	&	0.7002	&	3.99	&	8.12	\\
bdexp	&	B 5	&	0.0004	&	2.21	&	3.86	\\
pspdoc	&	B 5	&	4.9999	&	3.05	&	5.97	\\
augmlagn	&	$5\times5$ diagonal blocks	&	4.9998	&	4.15	&	9.28	\\
brybnd	&	B 11	&	12.9996	&	14.19	&	38.79	\\
chainros\_trigexp 	&	B 3 + D 6	&	4.4999	&	6.51	&	12.87	\\
toiqmerg 	&	B 7	&	6.9998	&	4.33	&	8.89	\\
arwhead	&	arrow	&	3.0000	&	3.63	&	6.78	\\
nondquar	&	arrow + B 3	&	4.9999	&	2.9	&	5.61	\\
sinquad	&	frame + diagonal	&	4.9999	&	5.12	&	10.01	\\
bdqrtic	&	arrow + B 7	&	8.9998	&	8.98	&	19.62	\\
noncvxu2	&	irregular	&	6.9998	&	4.95	&	9.55	\\
ncvxqp3 	&	irregular 	&	6.9997	&	2.9	&	6.48	\\
heavey\_band	&	B 39	&	38.9995	&	20.74	&	61.27	\\ \hline
\end{tabular}									
\caption{Description of problem set together with the execution time in seconds of \texttt{edge\_push} and
\texttt{RevHedir} for $n=10^6$.}
\label{tab:tests}									
\end{table}									

We did not benchmark our results against an alternative algorithm for we could not find a known AD package that is capable of efficiently calculating such directional derivatives for such high dimensions. For small dimensions, we used the \texttt{tensor\_eval} of ADOL-C to calculate the entire tensor using univariate forward Taylor series propagation\cite{Griewank2000}. Then we contract the resulting tensor with the vector $d$. This was useful to check that our implementation was correct, though it would struggle with dimensions over $n=100$, thus not an appropriate comparison. 

A remarkable feature of these tests, is that the time spent by \texttt{RevHedir} to calculate $D^3f(x)\cdot d$ was, on average, 108\% that of calculating $D^2f(x).$ Thus, if the user is prepared to pay the price for calculating the
Hessian, he could also gain some third order information for approximately the same cost. The code for these tests can
be downloaded from the Edinburgh Research Group in Optimization website:
\href{http://www.maths.ed.ac.uk/ERGO/}{http://www.maths.ed.ac.uk/ERGO/}.

\section{Conclusion}
\label{sec:Conclusion}


Our contribution boils down to a framework for designing high order reverse methods, and an efficient implementation of the directional derivative of the Hessian called \texttt{RevHedir}. The framework paves the way to obtaining a reverse
method for all orders once and for all. Such an achievement could cause a paradigm shift in numerical method design,
wherein, instead of increasing the number of steps or the mesh size, increasing the order of local approximations
becomes conceivable. We have also shed light on existing AD methods, providing a concise proof of the
\texttt{edge\_pushing}~\cite{Gower:2012} and the reverse gradient directional derivative~\cite{Abate:1997} algorithms.

The novel algorithms~\ref{alg:RevHedir} and~\ref{alg:Reverse3rd} for calculating the third-order derivative and its
contraction with a vector, respectively, fulfils what we set out to achieve: they accumulate the desired derivative ``as
a whole'', thus taking advantage of overlapping calculations amongst individual components. This is in contrast with
what is currently being used, e.g., univariate Taylor expansions~\cite{Griewank2000} and repeated tangent/adjoint
operations~\cite{Naumann2012TAo}. These algorithms can also make use of the symmetry, as illustrated in our implementation of
\texttt{RevHedir} Algorithm~\ref{alg:Tdcomponent-wise}, wherein all operations are only carried out on a lower triangular
matrix.

We implemented and tested the \texttt{RevHedir} with two noteworthy results. The first is its capacity to solve sparse
problems of large dimension of up to a million variables. The second is how the time spent by \texttt{RevHedir} to
calculate the directional derivative $D^3f(x)\cdot d$ was very similar to that spent by \texttt{edge\_pushing} to
calculate the Hessian. We believe  this is true in general and plan on confirming this in future work through complexity
analysis.
Should this be confirmed,
it would have an immediate consequence in the context of nonlinear optimization, in that
the third-order Halley-Chebyshev methods could be used to solve large dimensional problems
 with an iteration cost proportional to that of Newton step. In more detail, at each step the Halley-Chebyshev methods require the Hessian matrix and its directional derivative. The descent direction is then calculated by solving the
Newton system, and an additional system with the same sparsity pattern as the Newton system.
If it is confirmed that solving these systems costs the same, in terms of complexity, then the cost of a
Halley-Chebyshev iteration will be proportional to that of a Newton step.
Though this comparison only holds if one uses these automatic differentiation procedures to calculate the derivatives in both
methods.
 
The CUTE functions used to test both \texttt{edge\_pushing} and \texttt{RevHedir} are rather limited, and further tests
on real-world problems should be carried out. Also, complexity bounds need to be developed for both algorithms.

A current limitation of reverse AD procedures, such as the ones we have presented, is their issue with memory usage. All floating point values of the intermediate variables must be recorded on a forward sweep and kept for use in the reverse sweep. This can be a very substantial amount of memory, and can be prohibitive for large-scale functions~\cite{Griewank2004}.
As an example, when we used dimensions of $n=10^7$, most of our above test cases exhausted the available memory on the personal laptop used. A possible solution to this, is to allow a trade off between run-time and memory usage by reversing only parts of the procedure at a time. This method is called checkpointing~\cite{Griewank2004,Sternberg2006}.

\bibliographystyle{siam}
\bibliography{Automatic_Differentiation}	
	
\end{document}